\documentclass{article}
\usepackage[T1]{fontenc}
\usepackage[utf8]{inputenc}
\usepackage{ismir}
\usepackage{amsmath,cite,url}
\usepackage{graphicx}
\usepackage{color}
\usepackage{gensymb}

\DeclareMathOperator*{\argmax}{arg\,max}
\usepackage{booktabs}
\usepackage{multirow}
\usepackage{makecell}

\title{Do Music Source Separation Models Preserve Spatial Information in Binaural Audio?}

\threeauthors
  {Richa Namballa} {New York University \\ \texttt{rn2214@nyu.edu}}
  {Agnieszka Roginska} {New York University \\ \texttt{ar137@nyu.edu}}
  {Magdalena Fuentes} {New York University \\ \texttt{mf3734@nyu.edu}}

\def\authorname{R. Namballa, A. Roginska, and M. Fuentes}

\usepackage[bookmarks=false,pdfauthor={\authorname},pdfsubject={\pdfsubject},hidelinks]{hyperref}

\begin{document}

\maketitle

\begin{abstract}

Binaural audio remains underexplored within the music information retrieval community. Motivated by the rising popularity of virtual and augmented reality experiences as well as potential applications to accessibility, we investigate how well existing music source separation (MSS) models perform on binaural audio. Although these models process two-channel inputs, it is unclear how effectively they retain spatial information. In this work, we evaluate how several popular MSS models preserve spatial information on both standard stereo and novel binaural datasets. Our binaural data is synthesized using stems from MUSDB18-HQ and open-source head-related transfer functions by positioning instrument sources randomly along the horizontal plane. We then assess the spatial quality of the separated stems using signal processing and interaural cue-based metrics. Our results show that stereo MSS models fail to preserve  the spatial information critical for maintaining the immersive quality of binaural audio, and that the degradation depends on model architecture as well as the target instrument. Finally, we highlight valuable opportunities for future work at the intersection of MSS and immersive audio.

\end{abstract}

\section{Introduction}\label{sec:introduction}

In recent years, immersive experiences have gained popularity in various forms of media such as video games, concerts, and movies. The shift to virtual and augmented reality (VR/AR) requires not only realistic visual stimuli, but authentic auditory cues as well. One common form of spatial audio used to provide the listener with directionality of sound is binaural audio. Binaural audio goes beyond traditional gain-based stereo panning by filtering two-channel audio to create interaural cues differing in level, time, and spectral content to simulate the location of a source in space~\cite{Roginska2017}. Furthermore, binaural audio requires reproduction through headphones or loudspeakers equipped with crosstalk cancellation to maintain spatial imaging integrity. Level differences resulting from the ``head-shadow effect'' and the Time Difference of Arrival (TDOA) of a sound at each ear provide directional cues. Frequency-dependent filtering, determined by the form of the listener's head and specific ear (pinna) shape, causes two identical sound sources positioned differently to exhibit slightly different spectral content at each ear, further assisting localization. The two common methods of producing binaural audio are recording with a binaural dummy head and signal processing with a Head-Related Transfer Function (HRTF). 

Beyond the increasing demand for immersive VR/AR experiences, binaural audio has significant potential applications in accessibility. For instance, individuals who identify as neuro-divergent or hard of hearing often benefit from enhanced auditory clarity, enabling them to isolate and focus on specific sound sources in complex acoustic environments, facilitating independent navigation and interaction in social and public settings. Binaural source separation has been shown to significantly enhance auditory accessibility by reducing background noise and emphasizing relevant auditory signals in real-time with the use of microphone-enabled headphones~\cite{Veluri2023}. In this context, music source separation (MSS) in binaural audio could substantially improve how individuals engage with and enjoy musical environments such as concerts, festivals, and other live performances, enabling users to isolate specific musical elements or instruments and thus enhance their listening experience and overall participation in music events. These tools can further be utilized for recorded binaural content such as spatial audio captures of live performances or binaural field recordings.

Despite these potential benefits and growing interest, binaural audio processing has received limited attention within the music information retrieval (MIR) community, particularly concerning MSS. In this work, we investigate whether existing MSS models are able to separate binaural mixtures into their respective stems while preserving the spatial characteristics, which are crucial for the immersive experience provided by binaural audio. We create a binaural MSS dataset based off of the well-established MUSDB18-HQ dataset~\cite{Rafii2019}, and leverage several  metrics that quantify separation quality, spatial distortion, and immersiveness to evaluate these models. Our results show that there is a considerable gap in binaural MSS performance compared with MSS in simpler stereo settings, and that this gap depends on model architecture and target source. Lastly, we discuss the shortcomings of current metrics and identify opportunities for future research.

\section{Related Work}\label{sec:related_work}

Until now, most work on binaural source separation has been completed in the speech domain, often overlapping with the similar task of target source extraction (TSE). In particular, the speech research community describes the task as two-fold: source separation and localization~\cite{Deleforge2012}. We focus on prior studies concerning the former.

Early two-channel source separation models were primarily signal processing-based, with a focus on mathematical and theoretical techniques~\cite{Torkkola1999}. As the focus moved towards capturing directionality, models began using psychoacoustic spatial cues to improve the performance of the signal processing-based source estimation methods~\cite{Van1988, Jourjine2000, Viste2003, Schulz2007, Kim2011, Abdipour2015}. With the technological progress made in computational resources, binaural source separation models shifted to using deep learning approaches to perform source extraction in more complex environments and in real-time~\cite{Zakeri2021, Yang2024, Zhang2017, Veluri2023TSE, Veluri2023}. Recent deep learning systems have proposed novel loss functions aimed at preserving the level, phase, and time differences between binaural channels, cues which are critical to the immersive nature of binaural audio~\cite{Hernandez2024, Tokala2024}.

To the best of our knowledge, the only published work on binaural MSS thus far concerns vocal separation of binaural audio recorded with a dummy head~\cite{Kasak2023}. Their approach uses various hybrid combinations of single- and multichannel-source separation algorithms to extract the vocal stems, with a focus on signal-processing methods~\cite{Rickard2007, Hyvarinen2000, Driedger2014, Seetharaman2017, Rafii2012}. The results are evaluated with standard source separation metrics~\cite{Vincent2006} and subjective listener ratings. Based on the limited existing research in binaural MSS, we believe that there is a significant opportunity to explore this task using deep learning methods, inspired by recent progress in the speech community.

Regarding performance, the most common metric reported for evaluating source separation models is the Signal to Distortion Ratio (SDR), measured in decibels (dB)~\cite{Vincent2006}. Specifically, for MSS, researchers often benchmark their models on the test set of MUSDB18-HQ and report the SDR both overall and by instrument type ~\cite{sdx}. SDR (and its scale-invariant version, SI-SDR~\cite{LeRoux2019}) aim to reflect what portion of the estimated stem corresponds to the reference stem versus any error introduced by interference from other instruments, noise, and artifacts. While SDR is well-established for evaluating mono and stereo tracks~\cite{Vincent2007}, it does not specify the amount of spatial error introduced between channels in the model's estimated output, which is essential for evaluating the quality of binaural source separation. Therefore, we leverage other metrics from the literature which reflect spatial quality.

In the immersive audio research community, there are several models used to quantify the quality of a binaural signal, such as BAM-Q~\cite{Flessner2017} and MoBi-Q~\cite{Flessner2014}, trained on a combination of extracted binaural features and subjective quality ratings. We save the use of these models for future work in binaural MSS and choose to focus on more accessible and interpretable metrics, further explained in Section~\ref{subsec:metrics}, which originate from the duplex theory of sound localization~\cite{Rayleigh1907}. This theory states that, along horizontal plane (0\degree elevation), humans use two auditory cues to localize the direction of a sound: the interaural time difference (ITD) and the interaural level difference (ILD). ITD refers to the difference in time of arrival, at each ear, of a sound emitted from a source. Generally, a sound will reach the ipsilateral (closest to the source) ear faster than the contralateral (farthest from the source) ear. Likewise, the ILD is the difference in a sound's intensity as it arrives at the ipsilateral and contralateral ears. Originally, it was believed that ILD was the primary cue used for high frequency signals while ITD was for low frequencies~\cite{Roginska2017}. However, recent studies have shown that broadband signals require a complex interaction of the ITD and ILD to effectively identify a sound's location~\cite{Bernstein2007}.

The work in \cite{Watcharasupat2024} leverages this duplex theory of localization to propose two energy-ratio metrics for spatial evaluation: Signal to Spatial Distortion Ratio (SSR) and Signal to Residual Distortion (SRR). These measures are interpreted similarly to SDR, with SSR intended as a substitute for the Image to Spatial Distortion Ratio (ISR), proposed by~\cite{Vincent2007}. The spatial error is computed by projecting the reference signal to the estimated signal and optimizing for relative changes in gain and delay. From these projections, we can separate the distortion in spatial information (spatial error) from errors such as interference in the estimated signal (residual error). The ratios of SSR and SRR are defined in Section~\ref{subsec:metrics}.

\section{Dataset}\label{sec:dataset}

To directly compare the performance of various MSS models on both stereo and binaural audio, we created a binaural version of MUSDB18-HQ~\cite{Rafii2019}. MUSDB18-HQ is the uncompressed, 22kHz-bandwidth version of MUSDB18~\cite{Rafii2017} containing full-length, mixed music tracks from primarily Western pop and rock genres as well as their respective stems separated into vocals, drums, bass, and ``other''. The training and test sets consist of 100 and 50 songs, respectively. All audio files are stereophonic in WAV format, sampled at 44.1kHz/16b. We call our binaural dataset Binaural-MUSDB and we refer to the original MUSDB18-HQ as Stereo-MUSDB.

\begin{figure}[ht]
  \centering
  \includegraphics[alt={Diagram illustrating the random placement of instrument sources in Binaural-MUSDB. Azimuth angles range from +90° (far left) to -90° (far right), with 0° at the center, directly in front of the listener. The figure visually represents the variability in spatial positioning used during dataset creation to simulate the distribution of sound sources in the horizontal plane.},width=0.9\linewidth]{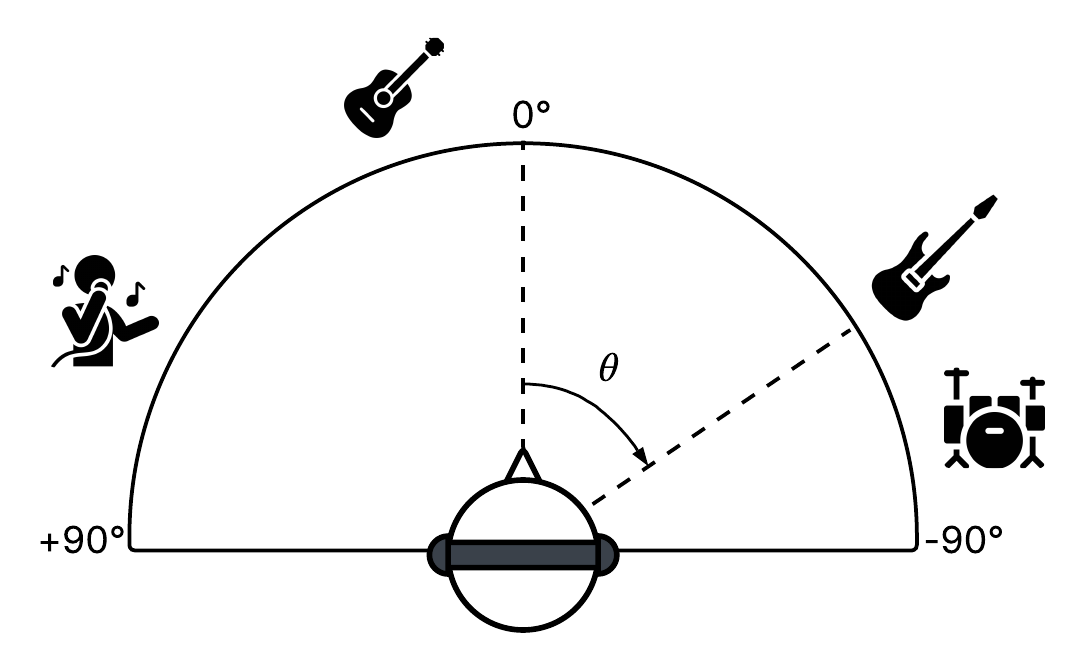}
  \caption{Binaural-MUSDB: each binaural source signal $\mathbf{s}_i$ is placed randomly along the horizontal plane at an angle $\theta_i \in[-90\degree, +90\degree]$ with the origin located directly in front of the listener. Every source has a minimum of 10\degree separation from the others, ensuring that there is no direct spatial overlap between stems.}
  \label{fig:source_placement}
\end{figure}

\begin{figure}[h!]
  \centering
  \includegraphics[alt={A stacked histogram showing the distribution of instrument source directions in the 50 song test set of Binaural-MUSDB. The x-axis represents azimuth angles from -90° to 90° in 30° increments with each bin of the histogram representing 10°. The y-axis shows count values from 0 to 14. Four instrument categories are displayed: vocals, bass, drums, and other. The distribution is not perfectly uniform, but the sources do not appear to favor any specific location in particular.},width=0.95\linewidth]{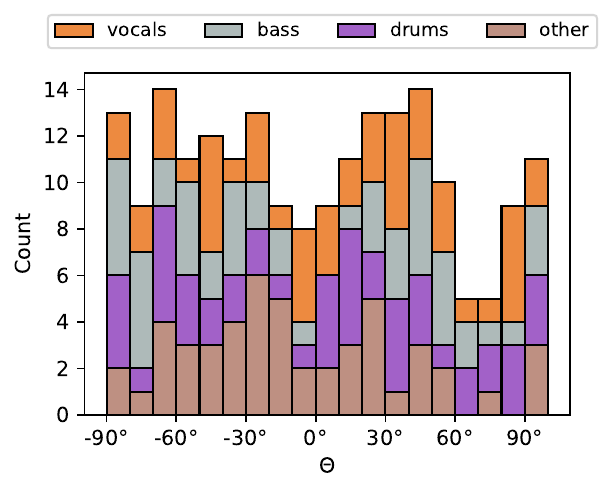}
  \vspace{-1.5em}
  \caption{Distribution of instrument positions in the test set of Binaural-MUSDB. $\theta$ corresponds to the source's location along the horizontal plane where 0\degree corresponds to the position directly in front of the listener.}
  \vspace{-1.5em}
  \label{fig:angles}
\end{figure}

To construct Binaural-MUSDB, we utilized binaural synthesis to create the illusion of the source signal emitting from a specific location around the listener~\cite{Roginska2017}. We use the publicly available SADIE II\footnote{\url{https://www.york.ac.uk/sadie-project/database.html}} database of HRTFs~\cite{Armstrong2018}. Each two-channel HRTF measurement contains the auditory spatial cues which can be superimposed onto a signal such that the listener will perceive the sound as originating from a location along the azimuth ($\theta$) and at a given elevation ($\phi$). For our synthesis, we apply the HRTF measurements for subject D1 from SADIE II, which correspond to the head and pinnae of the Neumann KU100 binaural dummy head microphone, which is the size of the average human head.

 We limited the horizontal plane to $\theta \in [-90\degree, +90\degree]$ along the azimuth, fixed at $\phi = 0\degree$ elevation. In spatial audio, $\theta = 0\degree$ corresponds to the location directly in front of the listener, equidistant from the left and right ears, as seen in Figure~\ref{fig:source_placement}. While the duplex theory states that humans primarily rely on ITD and ILD for binaural localization on the horizontal plane~\cite{Rayleigh1907}, they require spectral information for disambiguating front-back locations~\cite{Hofman1998}. Since we limit source locations to the front half of the sound field, we do not anticipate any significant differences in results using HRTFs other than the KU100's.

\begin{figure*}[ht]
  \centering
  \includegraphics[alt={A flowchart of the synthesis process used to create Binaural MUSDB. It combines the monophonic version of each stem with a head-related impulse response (HRIR) from the SADIE II database through convolution to generate the left and right channels of the binaural version of the stem.},width=0.85\linewidth]{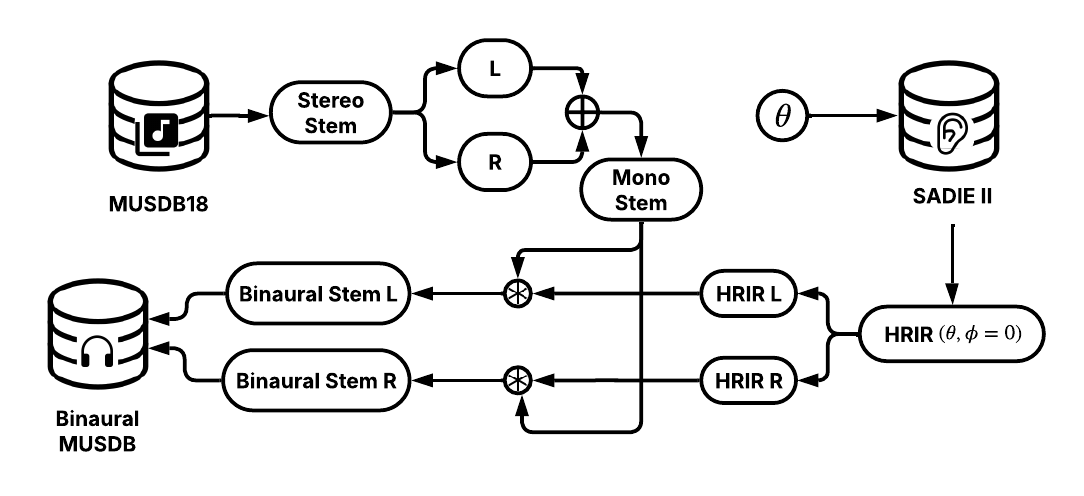}
  \vspace{-2em}
  \caption{An overview of the binaural synthesis process for the Binaural-MUSDB dataset. For every song in MUSDB18-HQ, each source is assigned a location $\theta$ along the azimuth in the frontal portion of the horizontal plane ($\pm90$\degree). The corresponding HRIR ($\theta$, elevation $\phi = 0$) is retrieved from the SADIE II database and each channel is convolved ($\ast$) with the monophonic version of the source stem. The resulting signals are the left and right channels of the binaural version of the stem which are included in the dataset.}
  \vspace{-1em}
  \label{fig:synthesis_process}
\end{figure*}

For every song in both the training and test sets, we assigned each source $i$ to a static location $\theta_i$ in increments of 10\degree. Angles for each stem in a single song were sampled randomly without replacement in the order of vocals, bass, drums, and other. Furthermore, in a given mixture, no two sources were allowed to be located at the same angle ensuring that there was a minimum of 10\degree separation (no direct overlap) between each stem. Each song was assigned only one set of source locations. The distribution of locations across the test set can be seen in Figure~\ref{fig:angles}.

We converted the original stereo stem to mono by averaging the two channels. Next, we loaded the Head-Related Impulse Response (HRIR), the time-domain version of an HRTF, corresponding to $\theta_i$ and convolved each HRIR channel with the mono stem signal to produce a binaural signal, with the two channels corresponding to the left and right ears. This process is visualized in Figure~\ref{fig:synthesis_process}. 

Finally, we summed the binaural versions of the vocals, drums, bass and other stems together and normalized the resulting signal to create the binaural mixtures which were used as the input to the MSS models described in Section~\ref{subsec:models}. The binaural synthesis was completed for all 150 tracks with the same train-test split as Stereo-MUSDB.

\section{Experimental Setup}

\subsection{Metrics}\label{subsec:metrics}

We utilize four metrics to describe the amount of distortion introduced by the MSS models, three of which quantify the level of spatial error in the estimated stems (SSR, $\Delta$ITD, $\Delta$ILD) and one that measures the remaining signal distortion due to interference and artifacts introduced by the separation (SRR).

In binaural audio, it is crucial that the ITD and ILD of a sound remain unchanged after separation to allow a listener to localize the source and maintain their sense of immersion. Therefore, we quantify how well the interaural cues are preserved by measuring the change ($\Delta$) in ITD and ILD between the estimated stem ($\mathbf{\hat{s}}$) and the reference stem ($\mathbf{s}$), as in~\cite{Veluri2023}. To compute $\Delta$ITD, we calculate the magnitude of the difference in ITD($\mathbf{\hat{s}}$) and ITD($\mathbf{s}$)~\cite{Han2020}. 

\vspace{-0.25em}
\begin{equation}\label{eq:delta_itd}
    \Delta\text{ITD} = |\text{ITD}(\mathbf{s}) - \text{ITD}(\hat{\mathbf{s}})|
\end{equation}

We measure the ITD of each signal as the TDOA of the source in the left and right channels using the frame-wise Generalized Cross Correlation with Phase Transform (GCC-PHAT) algorithm~\cite{Knapp1976}, implemented by~\cite{Veluri2023}. First, we segment the signal $\mathbf{x}$ into frames of 0.5s in length (with no overlap) and apply a Tukey window to each frame. Next, we calculate the GCC-PHAT $C(t, \tau)$ at frame $t$, for lags $\tau$ (in samples) corresponding to the range [-1, 1] ms, and find $\tau^*$, the value of $\tau$ which maximizes $C$~\cite{Veluri2023, May2010}. The frame-wise TDOA is computed in seconds by dividing $\tau^*$ by the sample rate $f_s$. 

\vspace{-0.25em}
\begin{equation}\label{eq:itd}
    \text{TDOA}(\mathbf{x}, t) = \frac{1}{f_s} \cdot \argmax_{\tau}\, C( t, \tau)
\end{equation}

The ITD of the full signal is then calculated as the weighted mode of the frame-wise TDOA. Each weight $w_t$ is based on the Root Mean Square (RMS) energy, where $x_{tc}$ is the signal at frame $t$ and channel $c$, $n$ is the length of the frame, and $k$ is the sample index of the frame.

\vspace{-0.25em}
\begin{equation}\label{eq:rms}
    w_t = \max_c \left( \sqrt{\frac{1}{n} \sum_{k=0}^{n-1}{x_{tc}[k]^2}} \right)
\end{equation}

Frames with a $w_t$ less than a threshold of $5 \times 10^{-4}$ are considered silent and excluded from the signal's ITD calculation. $\Delta$ITD is presented in microseconds ($\mu$s)~\cite{Veluri2023}.

The ILD is computed as the decibel ratio of the sum of squares for each channel across the entire signal. Here, $x_c$ represents channel $c$ of the full signal $\mathbf{x}$, $k$ is the corresponding sample index, and $N$ is the length of the entire signal in samples. As with ITD, we report $\Delta$ILD.

\vspace{-0.25em}
\begin{equation}\label{eq:ild}
    \text{ILD}(\mathbf{x}) = 10 \cdot \log_{10} \left(\frac{ \sum_{k=0}^{N-1}{x_L[k]^2}}{ \sum_{k=0}^{N-1}{x_R[k]^2}}\right)
\end{equation}

\vspace{-0.25em}
\begin{equation}\label{eq:delta_ild}
    \Delta\text{ILD} = |\text{ILD}(\mathbf{s}) - \text{ILD}(\hat{\mathbf{s}})|
\end{equation}

For both $\Delta$ITD and $\Delta$ILD, a lower value indicates a higher-quality spatial preservation of the interaural cue in the estimated stem.

In addition to $\Delta$ITD and $\Delta$ILD, we compute the SSR and SRR as proposed in~\cite{Watcharasupat2024} using their provided open-source implementation with its default parameters. Both metrics are computed frame-wise, reporting the median value, with a window of 1s and a hop length of 0.5s.

\vspace{-0.25em}
\begin{equation}\label{eq:ssr}
    \text{SSR}(\mathbf{\hat{s}}; \mathbf{s}) = 10 \cdot \log_{10} \left(\frac{||\mathbf{s}||^2}{||\mathbf{e}_{\text{spat}}||^2}\right)
\end{equation}

\begin{equation}\label{eq:srr}
    \text{SRR}(\mathbf{\hat{s}}; \mathbf{s}) = 10 \cdot \log_{10} \left(\frac{||\mathbf{\Tilde{s}}||^2}{||\mathbf{e}_{\text{resid}}||^2}\right)
\end{equation}

The SSR is intended to capture the spatial distortion introduced by the separation ($\mathbf{e}_\text{spat}$) into the estimated stem ($\mathbf{\hat{s}}$) while the SRR reflects only non-spatial distortion and errors such as interference and artifacts ($\mathbf{e}_\text{resid}$). Note that $\mathbf{\Tilde{s}}$ is the projection\footnote{Due to space constraints, we encourage readers to reference the original publication~\cite{Watcharasupat2024} for the precise mathematical definition of $\mathbf{\Tilde{s}}$.} of $\mathbf{s}$ into $\mathbf{\hat{s}}$, as mentioned in Section~\ref{sec:related_work}. Both SSR and SRR are measured in dB and a higher value indicates less distortion in the estimated signal.

\subsection{Models}\label{subsec:models}

We evaluate the performance of three well-known pre-trained MSS models on both stereo and binaural conditions: Hybrid Transformer Demucs Fine-Tuned (\texttt{htdemucs\_ft})~\cite{Rouard2023}, OpenUnmix (\texttt{umxhq})~\cite{Stoter2019}, and Spleeter (\texttt{spleeter:4stems})~\cite{Hennequin2020}. We chose these models over newer MSS models to validate our results with~\cite{Watcharasupat2024} and because all three models have official open-source implementations available for use. Both Demucs and OpenUnmix are trained on the Stereo-MUSDB training set, while Spleeter is trained on a proprietary dataset. Additionally, the version of Demucs we use is trained on an extra 800 songs not publicly identified. Each model accepts a stereophonic mixture input and returns an estimated two-channel stem.

Both OpenUnmix and Spleeter have inputs in the frequency domain, while Demucs is a hybrid model, operating in both the waveform and spectrogram domains. Spleeter uses a U-net architecture (CNN-based)~\cite{Ronneberger2015} to estimate a time-frequency mask for each source and applies it to the input mixture's magnitude spectrogram to generate the spectrogram of the estimated stem~\cite{Jansson2017}. OpenUnmix operates similarly, however, it uses a bi-directional LSTM model (RNN-based) to estimate the mask~\cite{Uhlich2017}. All three models use a L1 loss function to minimize the error between the estimated and reference signals.

To preserve the temporal structure of the input audio, both OpenUnmix and Spleeter apply the original input mixture's phase to the estimated magnitude spectrogram before inversion to the time domain to construct the final predicted stem. On the other hand, since Demucs functions in two domains, the model has to combine the estimated time and frequency representations to provide the final synthesized waveform. In the original hybrid version of Demucs~\cite{Defossez2021}, the model required careful hyperparameter tuning to align the temporal and spectral representations of the estimated signal so they could be summed in the waveform domain. However, in the newest version of the model~\cite{Rouard2023}, the authors claim that the transformer addresses this bottleneck through its flexible architecture.

To compare the separation performance in stereo and binaural settings, we apply these models to the test sets of Stereo-MUSDB and Binaural-MUSDB.

\section{Results and Discussion}

In this section, we analyze and discuss the performance of the three MSS models by looking at the different metrics in the binaural and stereo datasets, considering the effect on individual instruments, and identifying the effect of spatial distortion in the different locations along the azimuth.

\begin{table}[ht]
\centering
\vspace{-1em}
\caption{SRR results from the MSS models across the two datasets using median values. The best results are highlighted in {\bf bold} and the second best are \underline{underlined}.}
\vspace{0.3em}
\resizebox{\columnwidth}{!}{%
\begin{tabular}{@{}clccccc@{}}\toprule
\multirow{2}{*}{\bf Dataset} & \multirow{2}{*}{\bf Model} 
& \multicolumn{5}{c}{\bf SRR (dB) $\uparrow$} \\
\cmidrule(lr){3-7}
& & Bass & Drums & Other & Vocals & Overall \\
\midrule

\multirow{3}{*}{{\rotatebox[origin=c]{90}{\shortstack{Binaural}}}}
 & Demucs     & \textbf{8.90} & \textbf{10.58} & \underline{4.10} & \underline{4.37} & \underline{6.91} \\
 & OpenUnmix  & 3.37 & 6.75 & 1.19 & 2.37 & 3.51 \\
 & Spleeter   & 1.53 & 4.71 & 0.11 & 0.00 & 2.01 \\
\midrule

\multirow{3}{*}{{\rotatebox[origin=c]{90}{Stereo}}}
 & Demucs     & \underline{8.36} & \underline{9.86} & \textbf{6.36} & \textbf{6.08} & \textbf{7.39} \\
 & OpenUnmix  & 1.72 & 4.82 & 2.90 & 2.40 & 3.14\\
 & Spleeter   & 1.25 & 4.51 & 3.31 & 2.76 & 3.21\\
\bottomrule
\end{tabular}
}
\label{tab:srr}
\vspace{-1.5em}
\end{table}

\begin{table*}[ht]
\centering
\caption{Spatial metric results (SSR, $\Delta$ITD, $\Delta$ILD) from the MSS models for the two datasets using median values. The best results are highlighted in {\bf bold} and the second best are \underline{underlined}.}
\vspace{0.3em}
\resizebox{\textwidth}{!}{%
\begin{tabular}{@{}clcccccccccccccccc@{}}\toprule
\multirow{2}{*}{\bf Dataset} & \multirow{2}{*}{\bf Model} 
& \multicolumn{5}{c}{\bf SSR (dB) $\uparrow$} 
& \multicolumn{5}{c}{\bf $\Delta$ITD ${\mu}$s $\downarrow$} 
& \multicolumn{5}{c}{\bf $\Delta$ILD (dB)$\downarrow$} \\
\cmidrule(lr){3-7} \cmidrule(lr){8-12} \cmidrule(lr){13-17}
& & Bass & Drums & Other & Vocals & Overall 
  & Bass & Drums & Other & Vocals & Overall 
  & Bass & Drums & Other & Vocals & Overall \\
\midrule

\multirow{3}{*}{{\rotatebox[origin=c]{90}{\shortstack{Binaural\\MUSDB}}}}
 & Demucs    
   & 9.13 & 10.39 & \underline{12.62} & 8.70 & 10.59 
   & \underline{476.19} & \textbf{0.00} & \underline{22.68} & \textbf{0.00} & 68.03 
   & 0.20 & 0.31 & 0.57 & 0.42 & 0.39 \\ 

 & OpenUnmix 
   & \underline{10.94} & \underline{12.22} & 11.04 & 8.20 &  10.43
   & 521.54 & \textbf{0.00} & 226.76 & \textbf{0.00} & 90.7
   & 0.41 & 0.38 & 0.72 & 0.73 & 0.50 \\

 & Spleeter    
   & 10.63 & 11.86 & 9.96 & 5.22 &  9.86
   & 544.22 & \underline{22.68} & \underline{22.68} & \underline{22.68} & \underline{22.68}
   & 0.44 & 0.52 & 0.99 & 0.74 & 0.64 \\
\midrule

\multirow{3}{*}{{\rotatebox[origin=c]{90}{\shortstack{Stereo\\MUSDB}}}}
 & Demucs
   & \textbf{17.18} & \textbf{20.63} & \textbf{14.11} & \textbf{13.42} & \textbf{16.01}
   & \textbf{0.00} & \textbf{0.00} & \textbf{0.00} & \textbf{0.00} & \textbf{0.00}
   & \textbf{0.08} & \textbf{0.07} & \textbf{0.11} & \textbf{0.05} & \textbf{0.08} \\

 & OpenUnmix 
   & 9.74 & 12.12 & 10.09 & \underline{11.22} & 10.73
   & \textbf{0.00} & \textbf{0.00} & \textbf{0.00} & \textbf{0.00} & \textbf{0.00}
   &\underline{0.12} & 0.10 & 0.24 & \underline{0.08} & \underline{0.12} \\

 & Spleeter    
   & 8.69 & 11.54 & 11.31 & 10.18 & \underline{10.78}
   & \textbf{0.00} & \textbf{0.00} & \textbf{0.00} & \textbf{0.00} & \textbf{0.00}
   & 0.15 & \underline{0.08} & \underline{0.23} & 0.10 & \underline{0.12} \\
\bottomrule
\end{tabular}
}
\label{tab:ss_metrics}
\vspace{-1em}
\end{table*}

\subsection{Stereo vs. Binaural Performance} 

Based on the median SRR values shown in Table~\ref{tab:srr}, we observe a relatively consistent separation quality across the stereo and binaural datasets, suggesting that introducing spatial cues does not dramatically impact the ability of models to isolate instruments from one another. The SRR serves as a proxy of separation quality in spatial audio settings as it considers all residual distortions that are not spatial. Demucs appears to outperform the other two models in SRR for both datasets, which aligns with its original SDR-based ranking reported on the test set of Stereo-MUSDB~\cite{Defossez2021, Stoter2019, Hennequin2020}.

The median spatial metrics in Table~\ref{tab:ss_metrics} show that the MSS models introduce substantial spatial distortion when applied to binaural audio. For reference, SSR values around 10dB relate to noticeable spatial distortion, while values below that indicate severe spatial distortion, based on trends seen in other energy-ratio metrics~\cite{Watcharasupat2024, MUSB18_SOTA, Vincent2006}. Note that spatialization in stereo tracks traditionally uses gain-based panning, so a median $\Delta$ITD of 0$\mu$s is not unexpected. Upon closer inspection, a few $\Delta$ITD values were nonzero, indicating that some interchannel temporal distortion is introduced by the models, even in the stereo stems.

Demucs shows a considerable performance drop from stereo to binaural conditions, especially in SSR, compared to the other models. A plausible explanation is that, by operating directly on waveforms, Demucs implicitly learned stereo spatial cues based on amplitude differences and struggled to effectively interpret the subtler spectral information characteristic of binaural audio. In turn, Open-Unmix occasionally achieves superior results in binaural settings compared to stereo, likely due to its frequency-domain masking approach that preserves the original mixture's phase, inadvertently maintaining the spatial integrity. Similarly, Spleeter, also employing frequency-domain masking, demonstrates stable and sometimes improved performance on binaural audio, reinforcing that preserving the original phase of a mixture can be beneficial for spatial cue accuracy. Nevertheless, none of the models' binaural metrics match Demucs's stereo performance level, demonstrating considerable room for improvement in retaining binaural spatial cues.

\begin{figure}[h!]
  \centering
  \includegraphics[alt={Boxplots showing the spatial performance of three music source separation models (HT Demucs FT, OpenUnmix, and Spleeter) across six azimuth angle bins ranging from -90° to 90°. The three metrics shown are Signal to Spatial Distortion Ratio (SSR in decibels), the change in Interaural Time Difference (Delta ITD in microseconds), and the change in Interaural Level Difference (Delta ILD in decibels). SSR values appear evenly distributed across angles. Delta ITD displays a clear U shape with higher median values and more variance at more extreme values. Delta ILD distributions vary across angles, from 0 to 3 decibels, although Demucs appears to perform the best among the models at all angles.},width=0.95\linewidth]{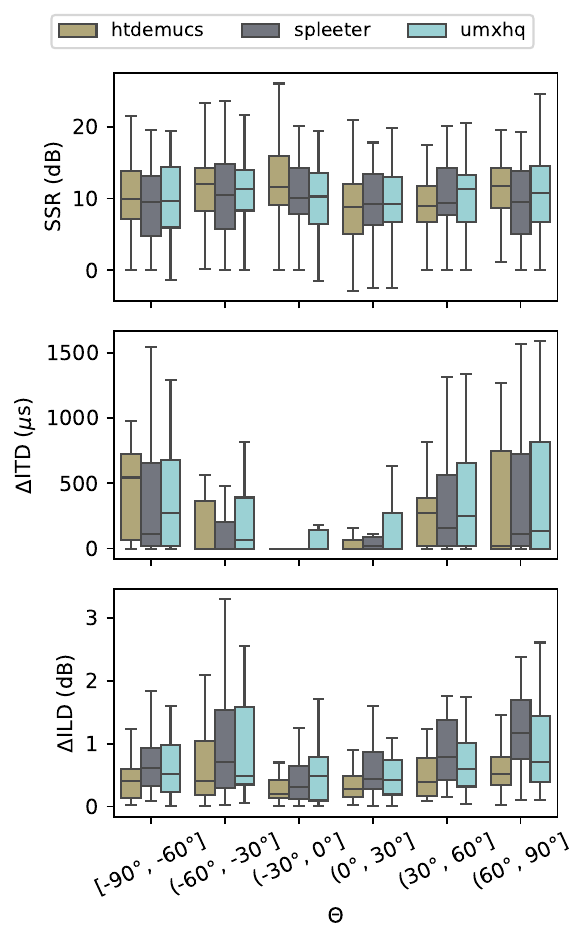}
  \vspace{-1em}
  \caption{Distributions of spatial metrics (SSR, $\Delta$ITD, $\Delta$ILD) by model and angle, aggregated across all sources.}
  \label{fig:metrics_angle}
  \vspace{-1.5em}
\end{figure}

\subsection{Performance by Angle}

Figure~\ref{fig:metrics_angle} shows the overall spatial distortion across all three spatial metrics by model and angle bin along the azimuth. We observe that SSR and $\Delta$ILD remain relatively consistent across angles, whereas the ITD notably distorts more the farther the source is positioned from the origin (larger $|\theta|$), displaying a U-shaped effect. One source of this tendency could be that strongly lateralized signals have minimal overlap in time-domain amplitude between the left and right channels. Cross-correlation relies on shared, correlated energy between channels so, in these cases, even minor disturbances from separation reduce channel similarity substantially, making accurate lag estimation challenging. This pattern could also imply that current MSS models are better at preserving amplitude-based spatial information (e.g., gain-based panning) than phase-based cues, and that they are introducing temporal disturbances. Additionally, the $\Delta$ITD distribution highlights a potential limitation in the SSR metric. Although it has been designed to account for all spatial distortions in accordance with the duplex theory~\cite{Rayleigh1907, Watcharasupat2024}, it may be more sensitive to level differences rather than time of arrival changes (as it does not reflect the U-shaped behavior observed in $\Delta$ITD). Further research with synthetic signals is needed to clarify how SSR values respond to phase distortions, whether the metric or its implementation requires revision, and how sensitive ITD and ILD calculations are to small artifacts.

\subsection{Performance by Instrument} 

When looking at instrument-specific performance in Tables \ref{tab:srr} and \ref{tab:ss_metrics}, we see that bass and ``other'' instruments exhibit higher spatial distortion ($\Delta$ITD) compared to vocals and drums. Bass instruments predominantly occupy narrow, low frequency bands, where localization relies heavily on subtle time differences rather than level. Because these low-frequency sounds have longer wavelengths, even minor phase distortions introduced during the separation process can lead to significant perceived spatial errors. This trait is reflected in the cross-correlation calculations of ITD, which require larger sample lags ($\tau$) to properly align the channels. Similarly, the ``other'' category often includes a diverse collection of complex and spectrally dense instruments with broader spatial positioning, resulting in diffused or ambiguous spatial cues.

\subsection{Performance by Model}

As mentioned previously, Demucs exhibits a significant performance drop from stereo to binaural conditions in terms of spatial distortion. In contrast, the frequency-domain models, Open-Unmix and Spleeter, display more consistent spatial performance across these two settings. Nevertheless, all models perform well below the level achieved by Demucs in stereo, suggesting that none are yet optimized for binaural spatial fidelity. Future research should explore training the models directly on binaural audio and adjusting the loss functions used during training to explicitly penalize distortions in ITD and ILD to improve spatial cue preservation, using systems inspired by the speech community~\cite{Veluri2023, Hernandez2024, Tokala2024}.

\subsection{Perceptual Considerations} While we primarily relied on objective metrics for our evaluation, preliminary subjective  listening by the authors suggests noticeable spatial distortions, particularly affecting bass instruments. These distortions align with our quantitative findings and indicate substantial spatial artifacts caused by inaccuracies in phase preservation. To provide a clearer illustration of these effects, selected audio examples demonstrating typical spatial distortions identified in our analysis are made available on an accompanying demonstration webpage, along with the open-source data and code repository.\footnote{\url{https://richa-namballa.github.io/binaural-mss-demo/}}

\section{Conclusion and Future Work}

We investigated the capabilities of existing music source separation (MSS) models applied to binaural audio. Our analysis revealed a considerable gap in MSS performance between binaural and stereo settings. This performance disparity was influenced significantly by both the specific architecture of the model and the target audio source. We identify several avenues of planned future work which will address the limitations of this study and build the foundation for subsequent binaural MSS models.

\textbf{Data.} The binaural data was synthesized with a random placement of sources and a single set of HRTF measurements. We hope to examine the stability of the results concerning the random seed initialization in the positioning of sources and the effect of their overlap. Additionally, we can validate the the impact of using diverse HRTFs (corresponding to various pinnae) when synthesizing the data.

\textbf{Metrics.} We believe the current metrics require further investigation to better understand their sensitivity to changes in phase versus level. Moreover, we can explore existing binaural quality models established by the immersive audio community and perform a perceptual study to validate all metrics.

\textbf{Modeling.} Since MSS research has progressed significantly, we hope to evaluate newer state-of-the-art MSS models' performance on binaural audio. We also plan to train a simple baseline MSS model on the binaural dataset with the option for data augmentations (e.g., noise, reverberation) to simulate diverse binaural conditions. Lastly, we will modify existing MSS model architectures to account for the preservation of spatial cues, such as with loss functions that minimize changes in ITD and ILD.

These paths for future research show promise in designing models specifically trained for binaural MSS with the goal of bridging immersive audio with music information retrieval for both cultural and accessibility applications.

\bibliography{references}

\end{document}